# Spin susceptibility of two dimensional hole gases in GaAs/AlGaAs heterostructures

Mukesh Kumar[a], Giorgio Mori[a, b,*], Flavio Capotondi[a, c], Giorgio Biasiol[a] and Lucia Sorba[a, c]

*a) Laboratorio Nazionale TASC-INFM, I-34012 Trieste, Italy.*

*b) Dipartimento di Fisica, Università degli Studi di Trieste, I-34127 Trieste, Italy.*

c) *Dipartimento di Fisica, Università degli Studi di Modena e Reggio Emilia, I-41100 Modena, Italy.*

## Abstract

We report measurements of spin susceptibility of a two-dimensional hole gas confined at an GaAs/AlGaAs interface in density range from 2.8 x $10^{10}$ to 6.5 x $10^{10}$ $cm^{-2}$. We used the method of spin polarization of carriers in parallel magnetic field. We observed an enhancement of the spin susceptibility over the bulk value that increases as the density is decreased in qualitative agreement with Quantum Monte Carlo calculations. The measured enhancement factor increases from 2.8 to 4.8 as the hole density is decreased.

## Introduction

The two-dimensional hole gases (2DHG) provide an ideal system for the study of Coulomb interaction effects in low dimensional structures because the hole effective mass is significantly larger than that of electrons. The strength of these interactions is measured by the interaction parameter $r_s$, which is the ratio of the carrier-carrier interaction energy to the kinetic energy of carriers. Thus, although the Coulomb interaction between holes is similar to that of electrons, the kinetic energy is lower in hole systems in the same density regime, making interaction effects more important. As the density of carriers is lowered, these interactions becomes stronger. It has long been expected that these strong interactions lead to a host of correlated states like Wigner crystallization at low temperatures [1, 2].

According to Quantum Monte Carlo (QMC) calculations the spin susceptibility ($\chi$), which is proportional to the effective g factor ($g^*$), increases with lowering of density [3-5]. The g factor is directly related to semiconductor band parameters and can differ strongly from its value in vacuum due to spin-orbit coupling [6, 7]. In AlGaAs/GaAs 2D-systems, the effective g factor is related to several effects as the carrier interaction, the 2D-confinement, the penetration of the carrier wave function into the barrier material [8, 9], and it is characterized a rather complicated behavior [6, 7].

Experiments in a number of two-dimensional electron gases (2DEGs) have confirmed the rise of spin susceptibility with decreasing density [10-15]. In some of GaAs-based two dimensional (2D) systems, measurements of spin polarization in parallel magnetic field [16-19], suggested a decrease of spin susceptibility with decreasing carrier density, opposite to the theoretical predictions [3-5]. This anomalous behavior was explained by *Tutuc et al.* by taking into account the finite layer thickness of the 2D system, which in presence of the parallel component of magnetic field leads to a change of effective mass and $g^*$ [20, 21]. On the contrary

*Zhu et al.* pointed out that the nonmonotonic behavior of the spin susceptibility is associated to the polarization and charge dependence of $\chi$ [15]. Therefore parallel field measurements give a spin susceptibility when the carriers are completely polarized.

Spin susceptibility of 2DHG in GaAs-based systems has not been studied in detail and only a few reports are available in literature [18-19].

We report here measurements of spin susceptibility at carrier polarization P=1 as a function of density in a high mobility 2DHG confined at a GaAs/AlGaAs interface. We found that the spin susceptibility, in the studied range of $r_s$ (12-20), increases monotonically with lowering the hole density in qualitative agreement with the QMC calculations.

## Experimental details.

The $GaAs/Al_{.33}Ga_{.67}As$ heterostructure was grown by a Veeco Gen II solid source molecular beam epitaxy (MBE) system on semi-insulating (311)A GaAs substrate. The 2DHG is confined in a single $GaAs/Al_{0.33}Ga_{0.67}As$ heterostructure with a Si-delta doping layer placed at a distance of 60 nm from the interface. The substrate temperature was kept at 635 °C during growth. The $As_4$ flux was kept constant throughout the growth, to a value corresponding to an $As_4$/Ga beam flux ratio of 14. The epitaxial layered structure was processed into six arm Hall bars of 250 μm length and 60 μm width, aligned along the [-233] direction by conventional optical photolithography and wet chemical etching. An Al gate was deposited by thermal evaporation on the top of the active region of the Hall bar to control the hole density.

The low temperature measurements were carried out in pumped $^3He$ cryostat at a base temperature of 260 mK. Standard four terminal low frequency (20 Hz) lock-in technique was used with a excitation current of 100 nA.

## Results and Discussion.

The 2DHG density (p) and mobility (μ) as measured by Hall effect at zero gate voltage are $1.4 \times 10^{11}$ cm$^{-2}$ and $7.5 \times 10^{5}$ cm$^{2}$/Vsec, respectively. The variation of mobility, when density is varied from $1.4 \times 10^{11}$ cm$^{-2}$ to $2.8 \times 10^{10}$ cm$^{-2}$ by applying an external gate voltage is shown in Fig. 1 (squares). The dotted and the dashed-dotted curves are the calculated mobilities for scattering with homogenous background doping ($\mu_{BI}$) and remote impurity doping ($\mu_{RI}$), respectively, according to the theory of Ref. [22]. For $p < 4 \times 10^{10}$ cm$^{-2}$ the remote impurity scattering is dominating while homogenous background doping is dominant for $p > 4 \times 10^{10}$ cm$^{-2}$. The solid line is the resulting mobility obtained by applying Matthiessen's rule, and it fits well the experimental data. The background doping density in our sample, as obtained from model of Ref. [22], is of the order of $10^{14}$ cm$^{-3}$. The inset of Fig. 1 shows that p varies linearly with the applied gate voltage.

Fig. 2 shows the longitudinal ($\rho_{xx}$) and transverse magnetoresistance ($\rho_{xy}$) of the 2DHG as a function of the perpendicular magnetic field. Absence of the parallel conduction is confirmed by verifying that longitudinal resistance in the integer quantum hall regime drops to zero. Oscillations corresponding to fractional quantum hall states are also visible. Fast Fourier transform (not shown here) of $\rho_{xx}$ versus 1/B gives a carrier density of about $1.38 \times 10^{11}$ cm$^{-2}$, in agreement with that measured by Hall effect, confirming again the absence of parallel conducting channels.

The hole effective mass ($m_h$*) of the 2DHG was determined by measuring the change in Shubnikov-de Haas (SdH) oscillation amplitude in low magnetic fields, as a function of temperature. The SdH oscillation amplitude decreases with increasing temperatures as shown in Fig. 3 and its dependence on effective mass arises from the dependence on the Landau level energy spacing which scales inversely to the effective mass [23]. In order to estimate the hole effective mass, we use the Dingle formula, $\Delta R/R_o \sim \xi/\sinh\xi$, where $\Delta R/R_o$ is the normalized amplitude of the SdH oscillations, $\xi = 2\pi^2 k_B T/h\omega_c$ and $\omega_c = eB/m^*$ [16]. Therefore, if the change of the oscillation amplitude, in the low field regime, is plotted as a function of temperature at a given magnetic field, it is possible to determine the effective mass by using it as an adjustable parameter to fit the experimental data. Figure 4 shows the result of such experimental fit for $\nu = 6$ that is highest Landau level (at lowest magnetic field) with well developed SdH oscillation in the studied temperature range. The hole effective mass which gives the best fit is $0.41 \pm 0.04\ m_e$. This value of effective mass is in close agreement with the value reported in literature for similar system [24]. The measurement of effective mass was repeated for several densities (not shown) and found to be constant within the experimental error in the density range under study.

In a normal Fermi liquid, the spin susceptibility is calculated to be $\chi = d\Delta n/dB = g^*\mu_B\rho_{DOS}/2$, where $\rho_{DOS}$ is the density of states at the Fermi level, and $\Delta n = n^{\uparrow} - n^{\downarrow}$, is the difference between the spin up and spin down carrier densities. In a 2D system, $\rho_{DOS} = m^*/\pi h^2$, therefore $\chi = g^*m^*\mu_B/2\pi h^2$, where $\mu_B$ is the Bohr magneton [15]. We express it as a relative spin susceptibility $\chi/\chi_0 = m^*g^*/m_b g_b$, where $m_b = 0.41m_e$ and $g_b = 0.5$ [25] are the values of hole effective mass and g-factor in GaAs, respectively, and $\chi_0$ is the Pauli susceptibility determined by these values.

We followed the parallel-field method recently utilized by several groups to derive $\chi$ ($\propto m^*g^*$) from the full polarization condition of 2D systems [10-15]. This method involves the application of an external parallel magnetic field ($B_{\|}$) which causes a Zeeman splitting of the energy bands. This splitting induces a difference in the population of the spin up and spin down subbands, which leads to a net spin polarization of the system. If the splitting exceeds the Fermi energy of the system, all spins are aligned and the 2DHG becomes fully spin polarized. Assuming that g* is independent of the applied magnetic field, we can write the splitting between the spin up and spin down subbands as $E_z = g^*\mu_B B_{||}$ in such a way that the 2DHG becomes fully spin polarized at a field $B_p$, given by [16]

$$B_p = (h^2/2\pi\mu_B)/(p/m^*g^*). \quad (1)$$

For the determination of $B_p$, we measured the in-plane magnetoresistance (MR) for different hole densities. The MR curve taken for $p = 3.1 \times 10^{10}$ $cm^{-2}$ is shown in Fig. 5a (solid line). This curve exhibits a strong positive MR. The $B_{||}$ dependence of $\rho_{xx}$ exhibits different behaviors at low fields and at high fields. In the low field regime, the magnetoresistance is proportional to $\exp(B_{||}^2)$ (dashed line), while in the high field regime, the magnetoresistance dependence changes to a simple exponential, $\exp(B_{||})$ (dotted line). It has been shown that the magnetic field, at which the simple exponential dependence starts, corresponds to the full polarization of the 2D system and is known as the polarization field $B_p$ [16-18]. The MR curves taken for different densities from $2.8 \times 10^{10}$ to $6.5 \times 10^{10}$ $cm^{-2}$ are shown in Fig. 5b. For higher densities, $B_p$ lied above our available magnetic fields. The small vertical lines in Fig. 5a and 5b indicate the corresponding $B_p$ for each density. The dependence of $B_p$ on p is shown in Fig. 6 (squares). $B_p$, according to Eq. (1) decreases linearly with decreasing p, (solid line represents the best fit to the experimental data).

Using the measured values of $B_p$ and the relation (1) spin susceptibility was determined. We plot the values of spin susceptibility normalized to the GaAs value $\chi/\chi_0$ in Fig. 7. The right hand side shows the value of g* calculated by assuming that the effective mass, constant with parallel magnetic field, is $m^* = 0.41m_e$. The measured values of spin susceptibility are from 2.8 to 4.8 times larger than the GaAs values and this enhancement increases with increasing $r_s$ (top abscissa axis of Fig. 7).

This behavior is in qualitative agreement with the theoretical calculations, which predicted that for an ideal 2D system the spin susceptibility should increase as $r_s$ increases and interactions become more important [4, 5]. Nevertheless Monte Carlo calculations predict a $\chi/\chi_0$ ratio 6 times higher [5]. In order to understand the discrepancy between our results and the theoretical calculations, several effects have to be analyzed.

In GaAs-based 2D electron systems, the decrease in spin susceptibility with lowering electron density can be explained by the finite layer thickness of the 2D system. In the presence of a large $B_{||}$, when the magnetic length, $l_B = (\hbar/eB_{||})^{1/2}$ becomes comparable to or smaller than the thickness of the 2D layer, the energy surface of the electrons gets deformed, which causes an increase of the effective mass [19, 20]. It is possible, as suggested in Ref. [19,20], that the same logic can also be applied to the 2D hole system like ours. Nevertheless in our sample, the spin susceptibility increases with decreasing hole density and it seems that the thickness of 2D layer is not playing any important role here. This idea is supported by the fact that the estimated background doping density in our sample is about $10^{14}$ $cm^{-3}$ which is higher than the one reported in Ref. [19]. Higher background doping decreases the thickness of 2D layer, while the 2D layer thickness increases with decreasing in carrier density [23]. For the above mentioned value of background doping and for the density range studied here, we calculated the 2DHG thickness using the model of Ref. [23]. It was found to vary approximately from 4 nm to 6 nm as p changes from the highest to the lowest value. On the other hand, the minimum value of magnetic length, which occurs at the maximum $B_p$, is about 8 nm. Since the 2DHG thickness in our sample is always less (at least 33%) than the magnetic length, application of a parallel magnetic field can not change the band structure and hole effective mass. Thus, effects related to finite layer thickness, mentioned in Ref. [19], should not playing a significant role in our sample.

The discrepancy between our data and the theoretical previsions could be also related to the presence of disorder in the hole system. At very low density, even a small amount of disorder may play a very important role. Disorder potential may lead to an inhomogeneous spatial distribution of the holes in the sample. If the holes are localized in potential minima, it is possible that the behavior of the 2DHG reverts to that of single particle system and hence g* decreases [16].

Another possible explanation could be related to the presence of strong spin-orbit interactions typical of GaAs-based systems. Since $g_b$ is significantly influenced by the spin-orbit coupling, it is expected that these interactions should also modify g* in a many-body picture [6,7].

In order to have a clear understanding of the mechanisms involved in the quantitative difference between theoretical and experimental measured $\chi/\chi_0$, further investigations are required.

## Conclusions

We have measured the spin susceptibility of a completely polarized 2DHG confined in a GaAs/$Al_{0.33}Ga_{0.67}As$ heterostructure. We found that the spin susceptibility qualitatively follows the theoretical predictions although the measured values $\chi$ are up to 6 times smaller. It seems that the presence of high background impurities in our sample limits the thickness of 2D layer to a value smaller than the magnetic length in the whole range of $B_p$ and p. Therefore the reduction of the spin susceptibility can not be ascribed to finite layer thickness of the system. The quantitative difference between experimental data and theoretical calculations may be due to disorder induced localization or/and spin-orbit interactions. Further studies are required to better understand the role of these effects on g* and hence on $\chi$ in GaAs-based 2D hole systems.

## Acknowledgements

We gratefully thank Prof. Gaetano Senatore, for fruitful discussions and suggestions. The first author (M. K.) would like to acknowledge TRIL program of Abdus Salam International Center for Theoretical Physics (ICTP), Trieste, Italy for providing financial assistance for carrying out the present research work. This work was supported in part by the CRS NEST-INFM in Pisa (Italy).

## References

* Corresponding author: Tel.: +39 040 3756439; fax: +39 040 226767.

*E-mail address:* mori@tasc.infm.it (G. Mori)

## Figure Captions

FIG. 1. Mobility versus hole density in our AlGaAs/GaAs heterostructure (squares). The dotted and the dashed-dotted curves are fits to the data for homogenous background doping ($\mu_{BI}$) and for remote impurity doping ($\mu_{RI}$), respectively. Solid line shows the resulting mobility. The inset shows the variation of p versus gate voltage.

FIG. 2. Perpendicular and parallel magnetoresistance as a function of perpendicular magnetic field at T= 260 mK.

FIG. 3. Parallel magnetoresistance as a function of perpendicular magnetic field at various temperatures.

FIG. 4. Stars: normalized Shubnikov-de Haas amplitude versus temperature for $\nu = 6$. Solid line: fits of the experimental data using the Dingle formula.

FIG. 5a. Measured magnetoresistance versus $B_{||}$ (solid line) at a fixed density, $p=3.1 \times 10^{10}$ $cm^{-2}$. The vertical line marks the polarization field $B_p$. In the low field regime the magnetoresistance dependence is proportional to $\exp(B_{||}^2)$ (dashed line), while in the high field regime the magnetoresistance dependence is proportional to $\exp(B_{||})$ (dotted line).

Fig 5b. In-plane magnetoresistance versus $B_{||}$ at different hole densities, from top to bottom: 2.9, 3.1, 3.4, 3.7, 4.3, 4.5, 6.1, and $7.4 \times 10^{10}$ $cm^{-2}$. Vertical lines mark $B_p$ for each density above which the magnetoresistance exhibits a simple exponential dependence.

FIG. 6. Squares: measured $B_p$ as a function of the hole density. The solid line is a linear fit of the data.

FIG. 7. Normalized spin susceptibility (squares) as a function of hole density and $r_s$. The right scale corresponds to g* assuming that the effective mass (m* = 0.41) does not change with $B_{||}$.

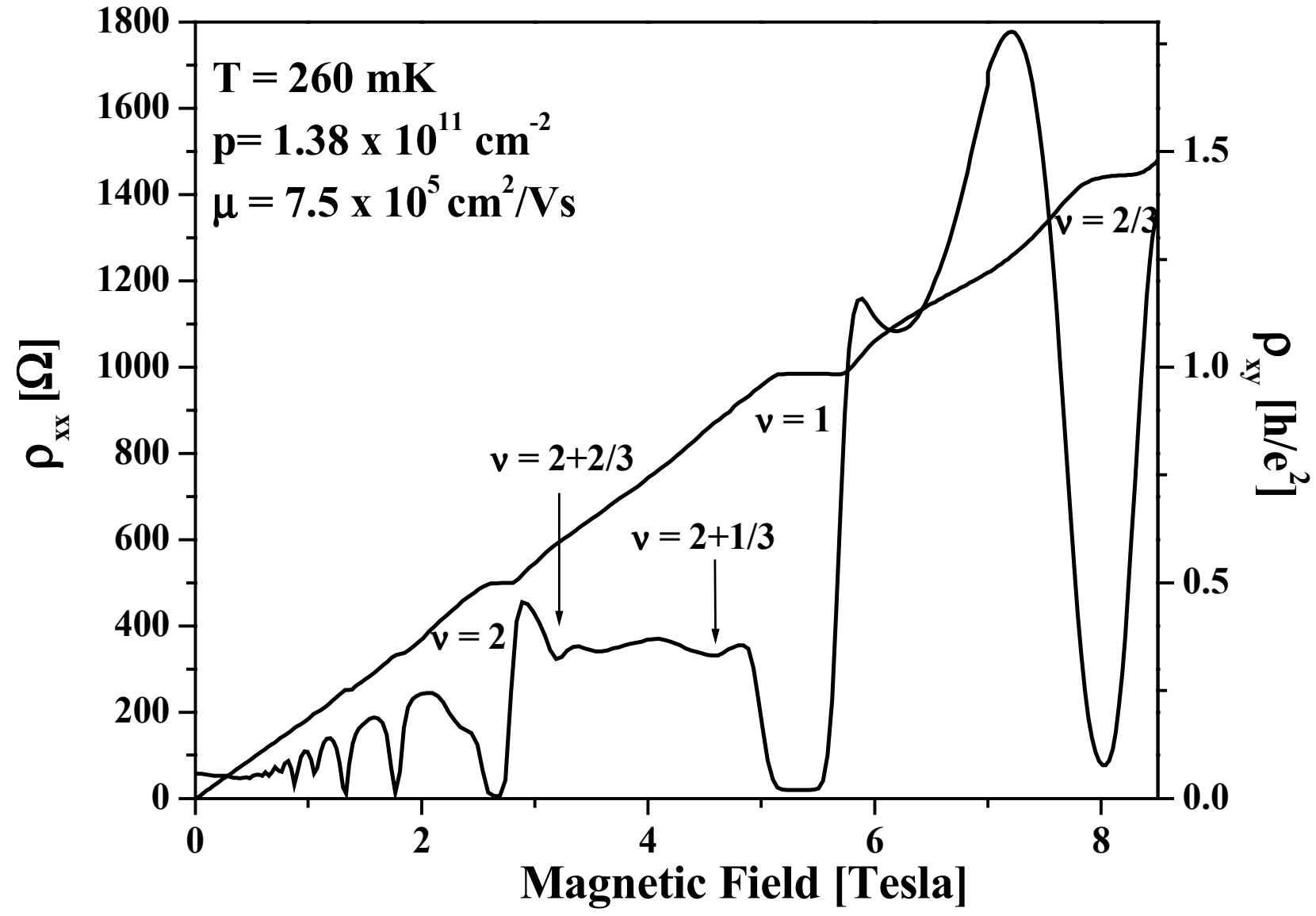

**Fig. 1.**

**M. Kumar et al.**

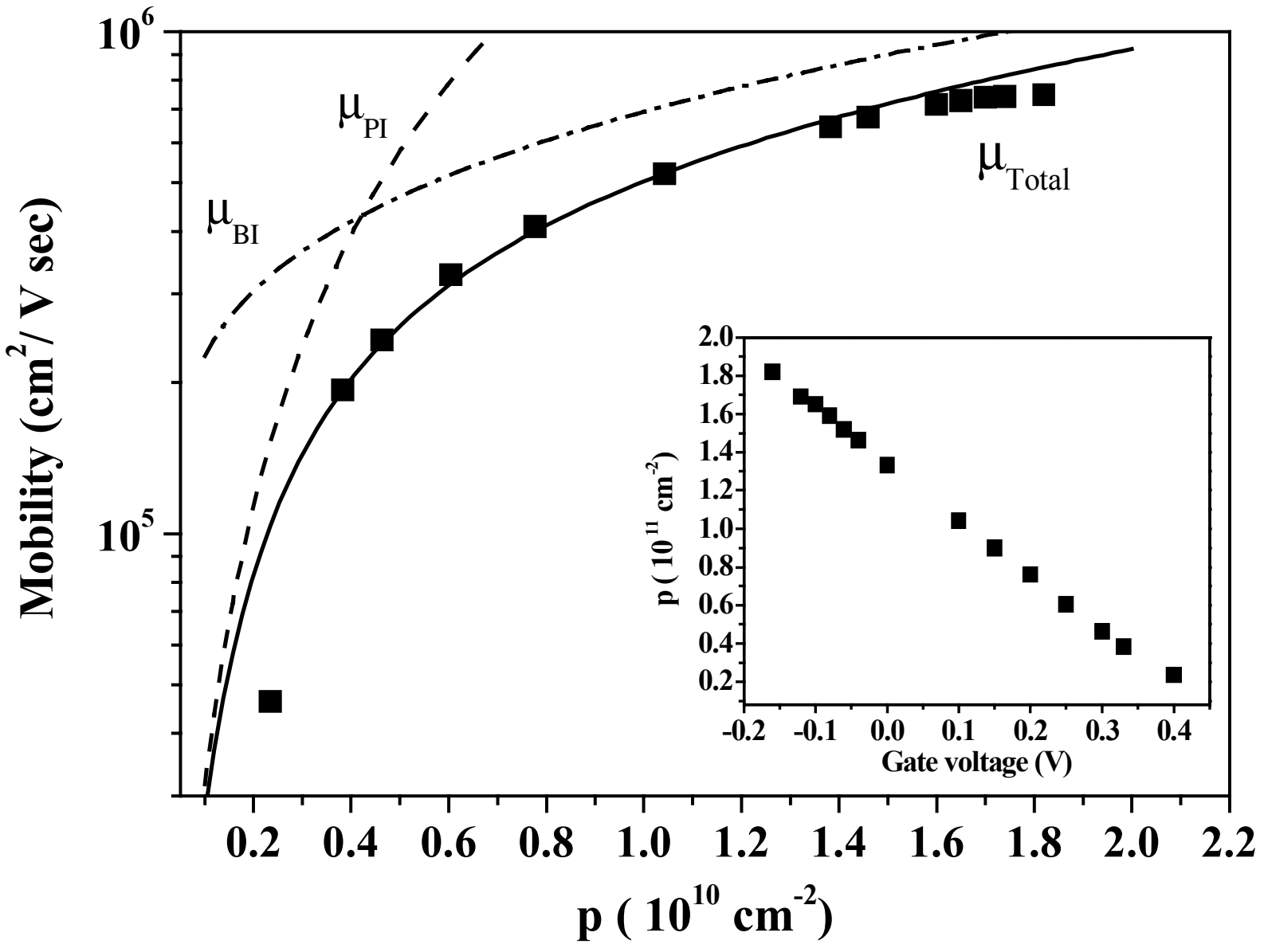

**Fig. 2**

**M. Kumar et al.**

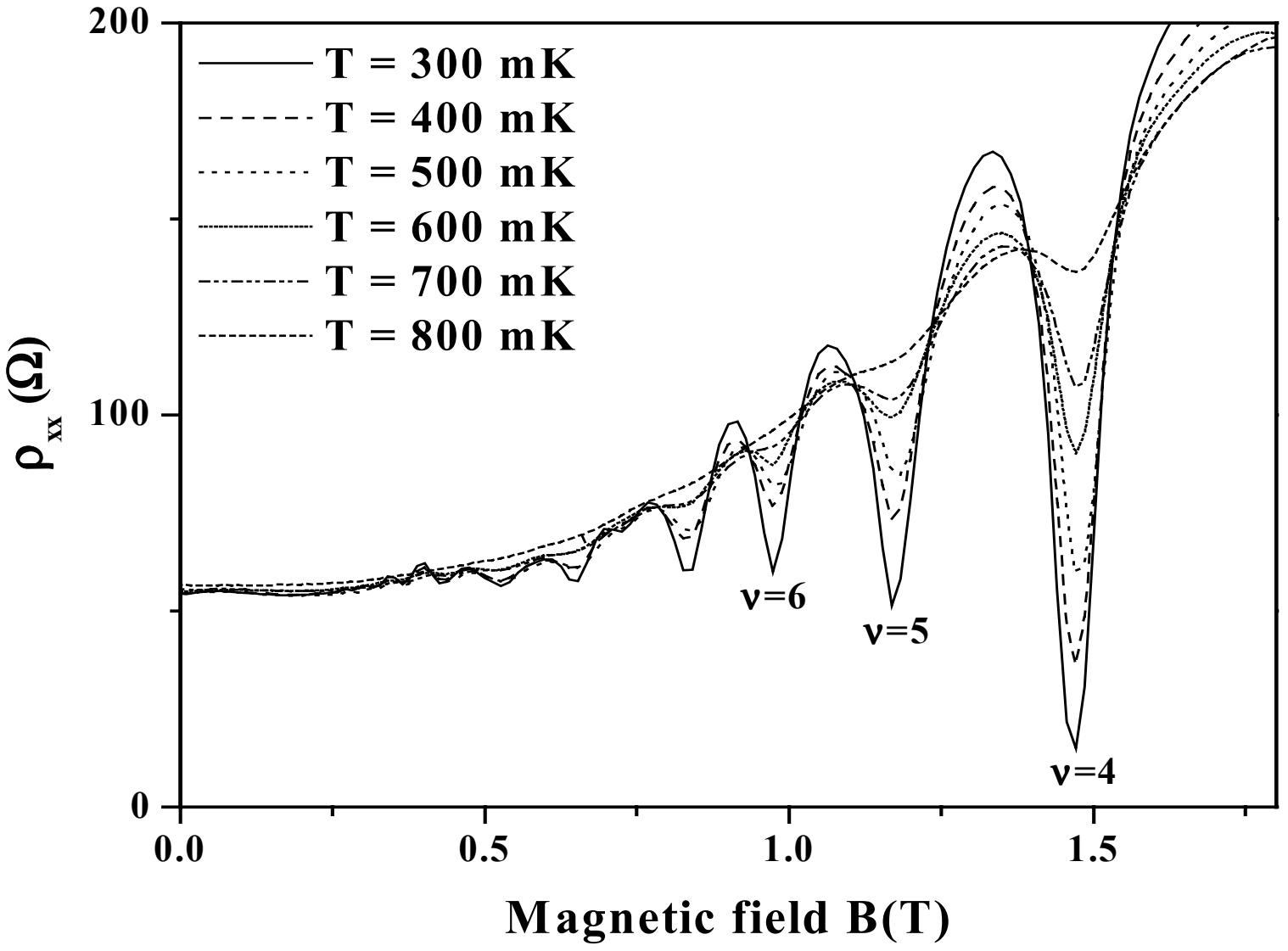

**Fig. 3.**

**M. Kumar et al.**

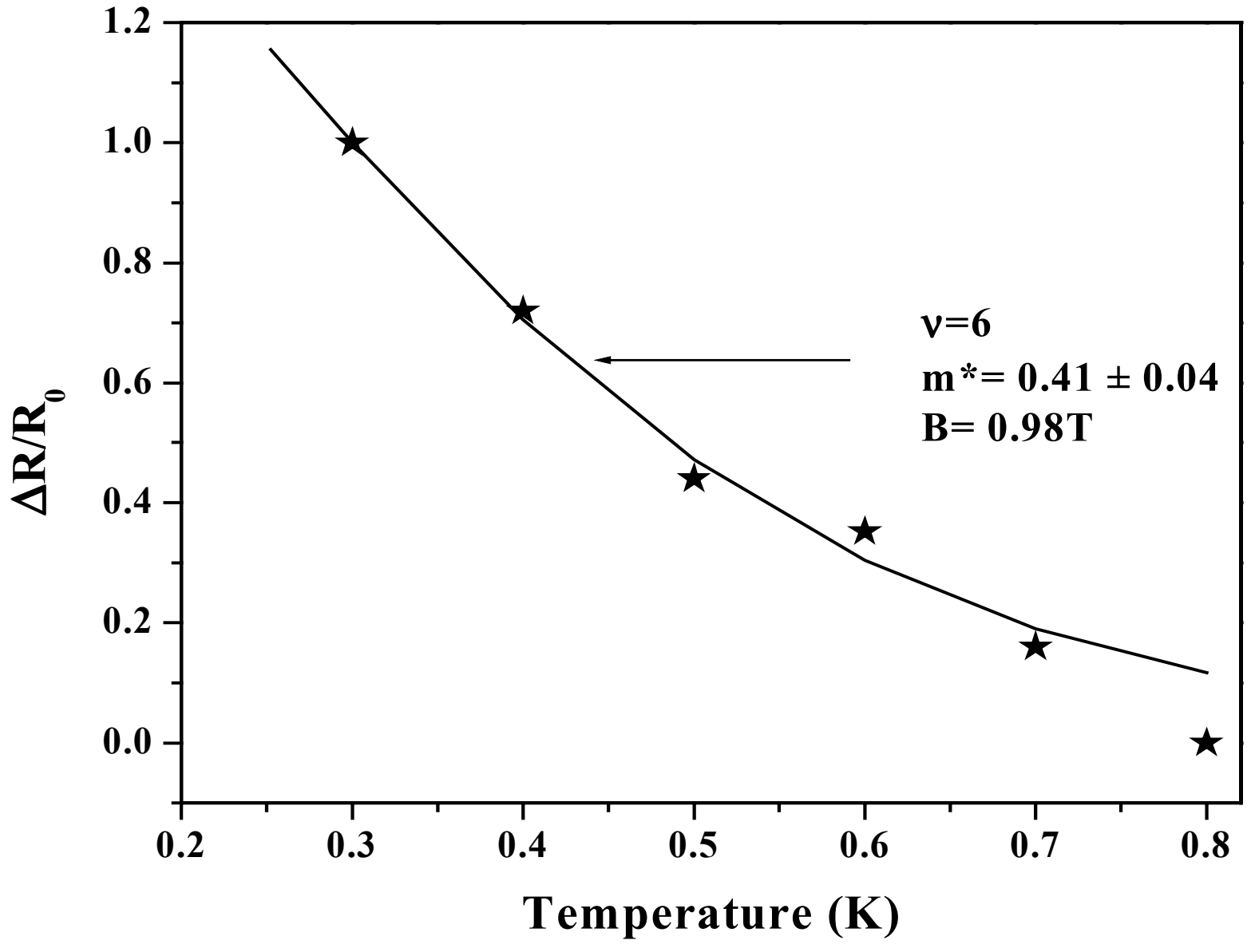

**Fig. 4.**

**M. Kumar et al.**

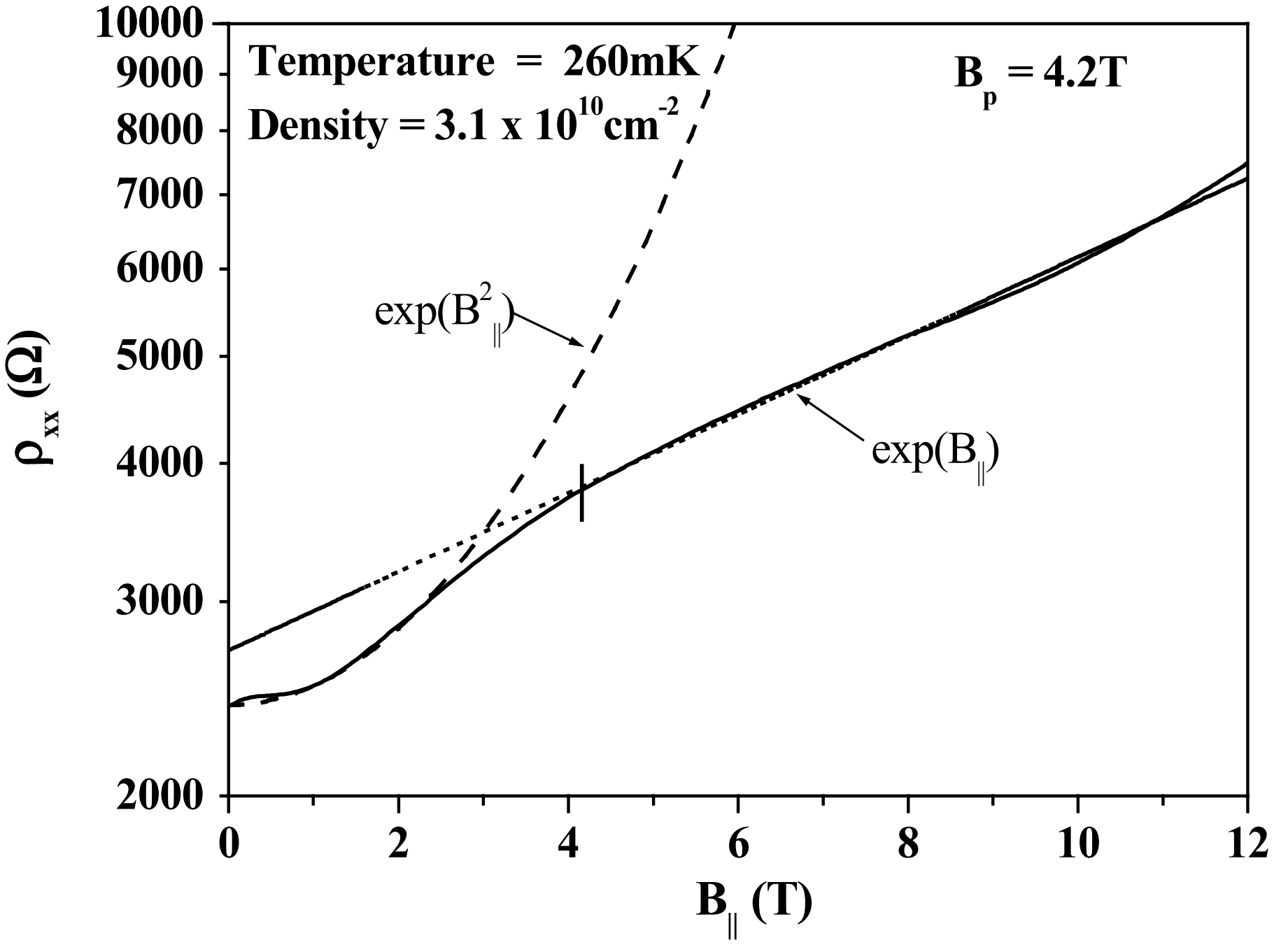

**Fig. 5a.**

**M. Kumar et al.**

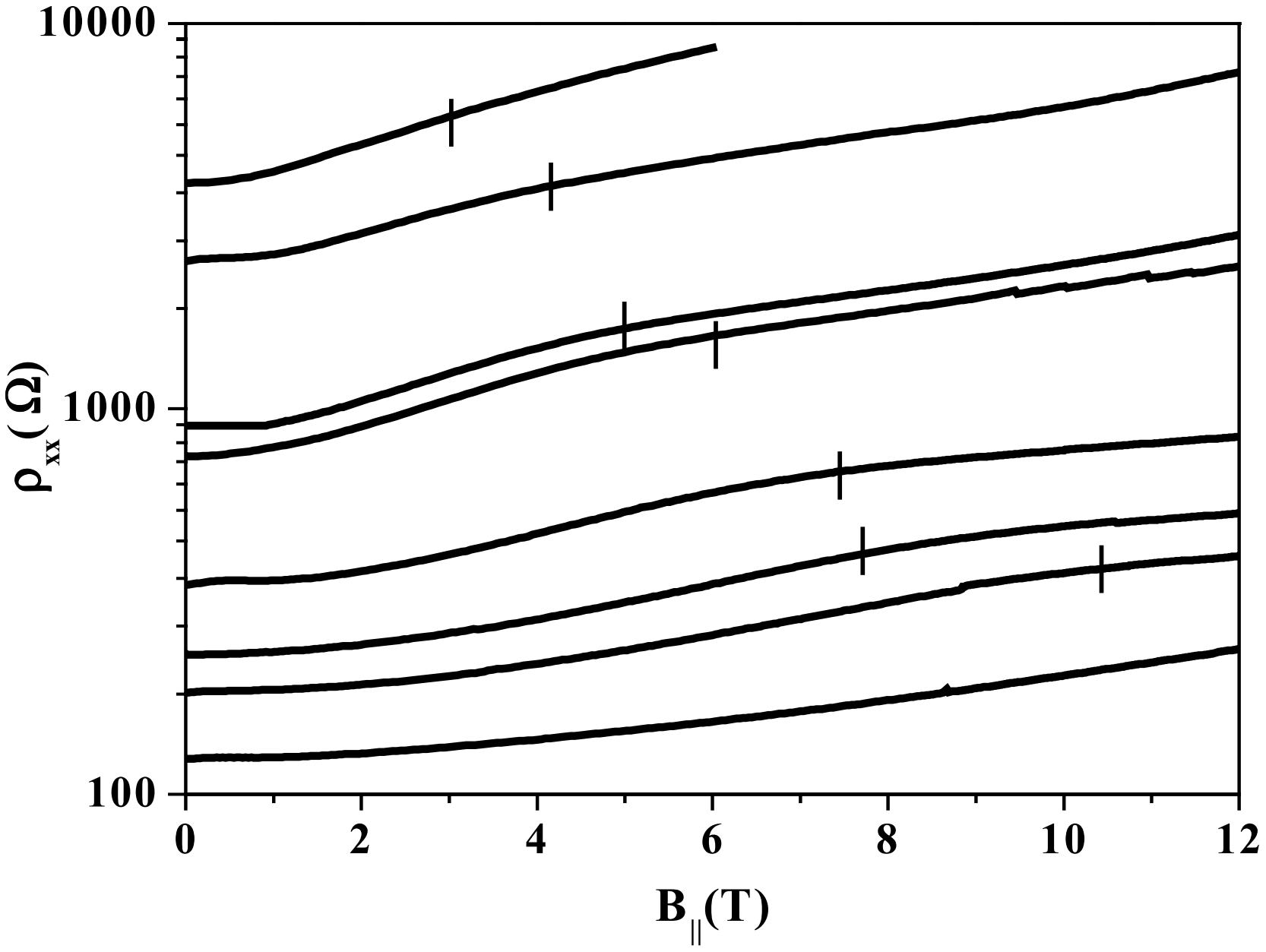

**Fig. 5b.**

**M. Kumar et al.**

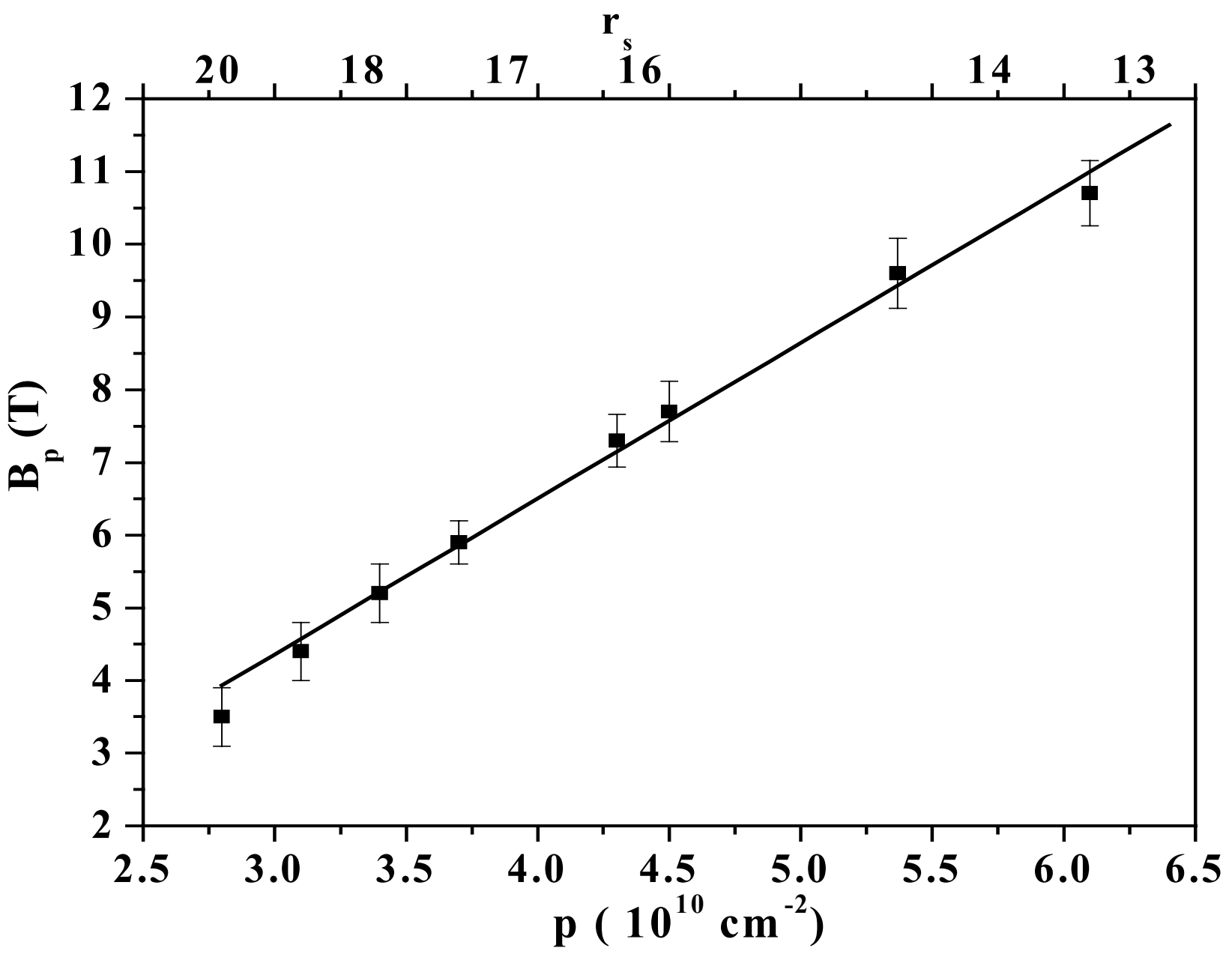

**Fig. 6.**

**M. Kumar et al.**

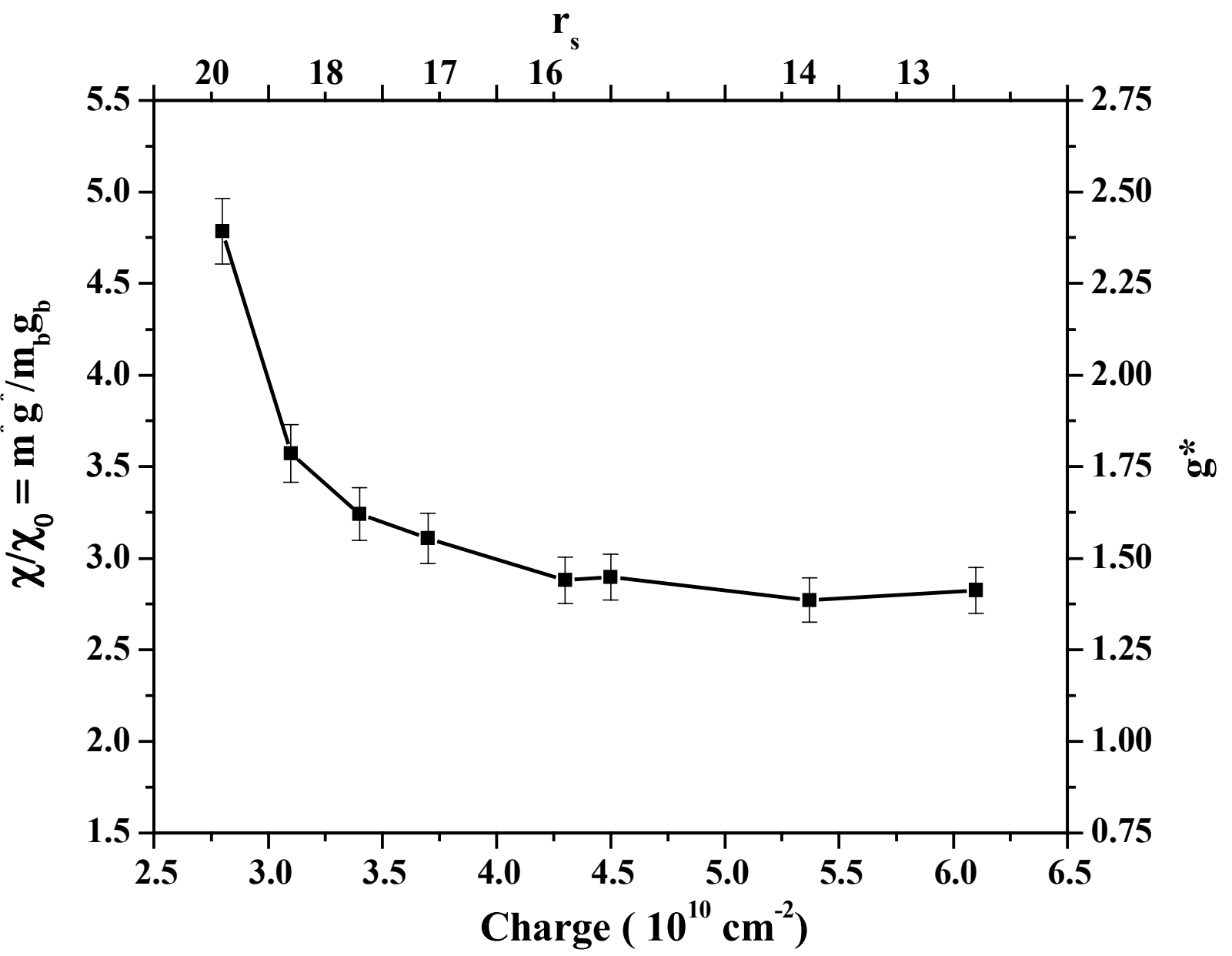

**Fig. 7.**

**M. Kumar et al.**